# Reconciling Shared versus Context-Specific Information in a Neural Network Model of Latent Causes

Qihong Lu[1*], Tan T. Nguyen[2], Qiong Zhang[3], Uri Hasson[1], Thomas L. Griffiths[1,4], Jeffrey M. Zacks[2], Samuel J. Gershman[5], Kenneth A. Norman[1]

1. Department of Psychology and Princeton Neuroscience Institute, Princeton University
2. Department of Psychological and Brain Sciences, Washington University in St. Louis
3. Department of Psychology and Department of Computer Science, Rutgers University-New Brunswick
4. Department of Computer Science, Princeton University
5. Department of Psychology and Center for Brain Science, Harvard University
*Correspondence and requests for materials should be addressed to qihong.lu@columbia.edu

Abstract

It has been proposed that, when processing a stream of events, humans divide their experiences in terms of inferred latent causes (LCs) to support context-dependent learning. However, when shared structure is present across contexts, it is still unclear how the "splitting" of LCs and learning of shared structure can be simultaneously achieved. Here, we present the Latent Cause Network (LCNet), a neural network model of LC inference. Through learning, it naturally stores structure that is shared across tasks in the network weights. Additionally, it represents context-specific structure using a context module, controlled by a Bayesian nonparametric inference algorithm, which assigns a unique context vector for each inferred LC. Across three simulations, we found that LCNet could 1) extract shared structure across LCs in a function learning task while avoiding catastrophic interference, 2) capture human data on curriculum effects in schema learning, and 3) infer the underlying event structure when processing naturalistic videos of daily events. Overall, these results demonstrate a computationally feasible approach to reconciling shared structure and context-specific structure in a model of LCs that is scalable from laboratory experiment settings to naturalistic settings.

In naturalistic settings, the stream of experiences is often not independent and identically distributed (iid). In particular, humans are constantly confronted with experiences that are context-dependent, and the underlying context is often temporally persistent. Consider the example of learning two similar languages, such as Norwegian and Swedish. First, natural language inputs are temporally auto-correlated – as a Norwegian travels to Sweden, he or she will experience Swedish for an extended period of time. As experiences are governed by auto-correlated contexts (Sweden versus Norway), learning can be facilitated by partitioning experiences according to their underlying contexts (or tasks) to prevent inter-context interference. In this example, lumping the two contexts, Sweden versus Norway, can cause interference for context-dependent responding (e.g., the word "water" is "vann" in Norwegian and "vatten" in Swedish). However, as many of the words are shared across these two languages (the word "book" is "bok" in both Norwegian and Swedish), it is also beneficial to dedicate a set of shared representations to maintain the regularities common across contexts. How should the brain maintain shared structure and context-specific structure simultaneously?

One promising approach to context-dependent learning is the theory of latent cause inference (LCI). This theory proposes that humans assign latent causes (LCs) to experiences to facilitate learning, prediction, memory, and generalization in a context-dependent manner (Figure 1)[1–7]. In the domain of event cognition, this theory was recently instantiated as a computational modeling framework known as Structured Event Memory (SEM)[3]. SEM addresses several important problems. First, SEM uses a Bayesian nonparametric algorithm to assign LCs to the observations. As the number of inferred LCs grows in an adaptive manner, SEM does not need to assume the number of contexts is known a priori. Second, SEM uses a dedicated neural network to represent each LC[3], which is related to the mixtures of experts architecture[8]. As representations of different LCs are fully separated into different networks, this circumvents catastrophic interference even if experiences are presented according to a blocked curriculum (i.e., a temporally extended period of exposure to experiences from one context, followed by a temporally extended period of exposure to experiences from a different context) – this is a key advantage of SEM over standard neural networks lacking this LC inference mechanism, which can show catastrophic interference in this situation[9] (see the discussion section for a summary of other relevant models of context-dependent learning).

The SEM framework accounts for a wide range of human data. First, SEM captured recent human data where blocked curricula facilitated learning of simple event schemas[9] (see Simulation 2). In this case, having separated context representations was important in reducing catastrophic interference. Second, SEM was able to explain the way humans divide up continuous streams of experiences into discrete events, a process known as event segmentation[10–15] (see Simulation 3).

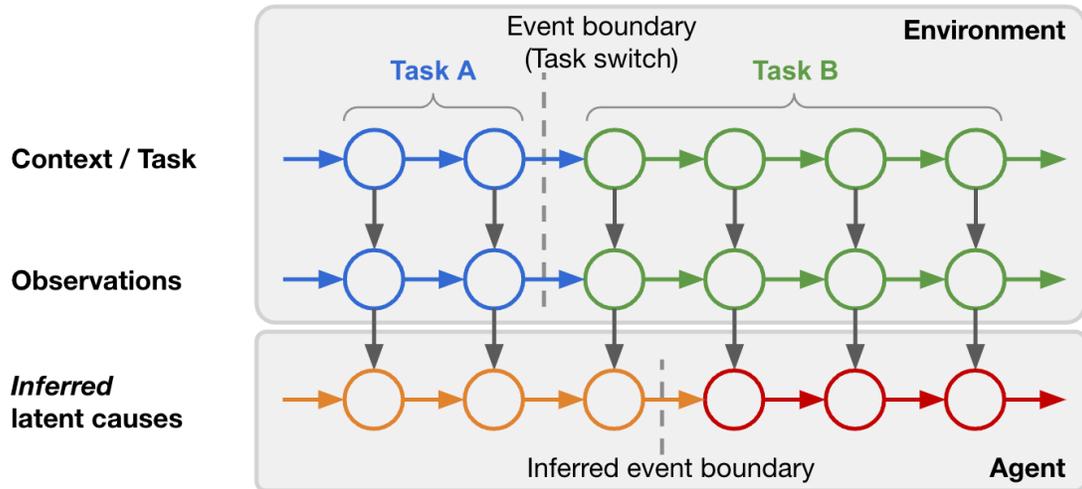

Figure 1. How latent cause inference partitions experience

Observations are governed by the current task identity, which is auto-correlated in time. At time t, the agent infers the latent cause (LC) of the current observation. Event segmentation is triggered when there is a switch of the inferred LC. Note that the inferred LCs do not always match the ground truth. In this work, the terms "context" and "task" are used interchangeably. Note that this is a conceptual diagram to clarify the terminology rather than a formal probabilistic model diagram.

---

While SEM has its strengths, it also has important shortcomings. In particular, its use of distinct representations for different LCs prevents it from representing shared structure. Because observations assigned to a particular LC are only used to train that assigned LC, other LCs do not learn at all from these observations. As such, SEM does not provide an answer to the question, raised above, of how we simultaneously represent context-dependent knowledge and shared structure. Moreover, when using fully separated neural networks to store knowledge about different contexts, the amount of required neural resource scales linearly as a function of the number of inferred LCs, which is computationally infeasible. The goal of this paper is to assess whether it is possible to modify the SEM architecture to better represent shared structure, without compromising its ability to account for human event segmentation (as evidenced by the two findings noted above). We know that the brain can do this, and it is a fundamental challenge for cognitive modeling to identify architectures that can jointly satisfy these two objectives. In the next section, we present a model as a step toward addressing this challenge.

## LCNet – Indexing latent causes with context vectors

We propose an alternative model, the Latent Cause Network (LCNet), which indexes LCs using context vectors (Figure 2b, 3c, 4c). This is inspired by the classic connectionist idea of context[16,17]. At a high level, instead of representing different LCs as different networks, LCNet represents different LCs in different regions in the hidden activity space of the same network. Similar to SEM, given an observation, LCNet decides whether to assign this observation to a known LC or a new LC using the same Bayesian LCI process, including the prior (see the Methods section for details of the latent cause inference procedure). Unlike SEM, in LCNet, each LC is represented by a unique random context vector sampled from a standard Gaussian distribution. There are two advantages of using random vectors: 1) new context vectors can be generated efficiently, and 2) random vectors in high dimensional space are approximately orthogonal (Supplement 1).

In Simulation 1, we provide a proof-of-concept simulation that LCNet can avoid catastrophic interference and also extract shared structure across tasks to make learning more efficient. Then, we show the ability to extract shared structure across tasks is fully compatible with two important sets of findings that SEM can explain. In Simulation 2, we show that, similar to SEM, LCNet can explain the finding showing that blocked curricula can facilitate learning of two simple event schemas[9], a result that is critically dependent on having separated representations of the two schemas. In Simulation 3, we show that LCNet can also explain human data on event segmentation while they view natural videos[18,19]. Overall, this set of results shows that it is possible to reconcile the tension between representing shared versus context-specific information.

Additionally, representing contexts as continuous vectors is a useful step towards a neurally plausible and computationally feasible context representation scheme; because LCNet uses context vectors to represent LCs instead of splitting off new networks, the size of the model no longer grows as a function of the number of inferred LCs – unlike SEM, inferring new latent causes does not lead to an increase in the number of parameters (weights) in the model.

## Simulation 1: Reusing prior knowledge of shared structure across tasks while reducing catastrophic interference

It is well known that humans can learn shared structure across tasks, making it possible to re-use prior knowledge more efficiently given a new task[20–24]. In this simulation, we present a proof-of-concept demonstration that LCNet can use the learned shared structure to learn new tasks faster, while overcoming catastrophic interference. Inspired by the literature on function learning[25–30], we designed a task set where the model had to learn multiple polynomial functions with shared structure – unbeknownst to the model, each function (Figure 2d) is a sum of an idiosyncratic term (Figure 2e) and a shared term (Figure 2f). As the shared term is shared across all tasks, learning the shared term should speed up the learning of new tasks. We performed this simulation $N$=100 times. Every time, all polynomial terms were re-sampled to make sure the results did not depend on the specific form of the polynomials.

Each polynomial function is a regression task. At time $t$, given the input value for the currently active task $x_t$ and a context-indicative signal (CIS) for the current polynomial, the model has to produce the corresponding output value $y_t$. Concretely, the CIS is simply a 128-dimensional random binary vector, sampled independently for every task. Noise is imposed by randomly flipping 10% of the values. The noise of the CIS was set to be low so that the LCI mechanism was guaranteed to be accurate for all models; this way, any differences in performance can be attributed to differences in the architectures being evaluated.

We compared three different models corresponding to three representation schemes for LCs: a standard feedforward neural network that uses fully shared representation (Figure 2a), a feedforward LCNet that uses random context vectors (Figure 2b), and a feedforward version of SEM that uses fully separated representations (Figure 2c). In feedforward LCNet, the same network is responsible for learning all polynomials, and that network receives context vectors that indicate which polynomial is currently active. In feedforward SEM, a unique network is dedicated to learning each polynomial. For both models, the LCI mechanism is responsible for inferring the ongoing LC. Finally, the regular feedforward model has to learn the task-dependent mapping with the CIS directly. See Algorithm 1 and Algorithm 2 in the methods section for a comparison of the differences between LCNet and SEM.

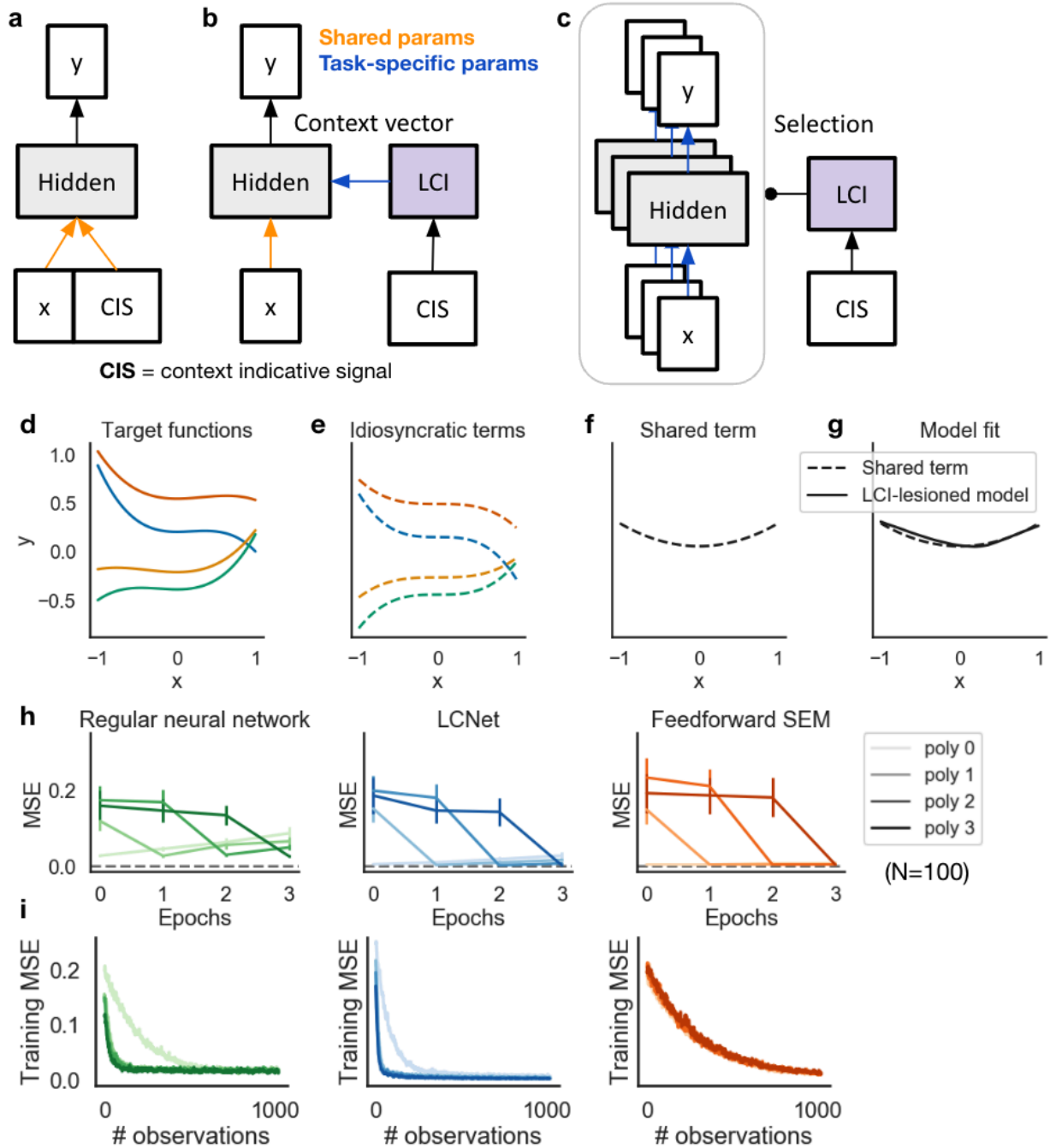

Figure 2. LCNet can overcome catastrophic interference and efficiently reuse the shared structure across tasks.

a) A standard feedforward neural network model. The weights are shared across all latent causes. b) In LCNet, the latent cause inference (LCI) mechanism infers the ongoing latent cause and feeds the corresponding context vector to modulate the activity of the hidden layer. Note that the neural network part is shared across tasks.

c) A feedforward SEM model[3], which uses K different neural networks to represent K latent causes. The LCI mechanism chooses which network to activate [3].

d) The polynomials that the models have to learn – each polynomial is a sum of the task-general shared term and a task-specific idiosyncratic term shown in panel e. These polynomials were resampled for every simulation run so that the results (h and i) were not specific to any particular choice of polynomials.

e) The task-specific terms used to generate the polynomials in panel d.

f) The shared term used to generate the polynomials in panel d.

g) The model produces the shared term when the LCI module is lesioned, which suggests that LCNet can factorize the task into a shared term and a task-specific term.

h) Learning curves for all polynomials plotted separately over epochs. The model was only trained on the i-th polynomial at epoch i. Error bars indicate 3SE. N = 100 models per condition.

i) Learning curves for each polynomial. Each curve is the learning curve within an epoch over the number of observations.

---

The results indicate that all models can learn to fit all four polynomials very well. To understand the role of LCI in LCNet, we lesioned the connection between LCI and the hidden layer (marked in blue, Figure 2b) after training, and we found that the output of the lesioned models produced the shared term, given the trained range of $x$ values (Figure 2g). This indicates that the model can factorize its knowledge in two different pathways – after training, the input-to-hidden pathway was mainly responsible for representing the shared term across all polynomials, while the LCI-to-hidden pathway was mainly responsible for the idiosyncratic term of the polynomials. The LCI-to-hidden pathway biases the final model output in a task-specific manner, and in the absence of it, the model produces what is common across tasks.

To examine if our model can 1) overcome catastrophic interference and 2) re-use prior knowledge efficiently, we looked at the learning curves, measured in terms of the mean squared error (MSE). All models were trained in a fully blocked setting where training was organized into distinct epochs. In epoch one, the model was only trained on randomly sampled observations from polynomial zero; in epoch two, the model was only trained on randomly sampled observations from polynomial one, and so on. Each observation is an $x$-$y$ pair such that the model had to make a regression prediction.

At the end of every epoch, all models were tested on all polynomials. The test MSE for the four polynomials was plotted separately, and the lines with lower color transparency correspond to polynomials that were trained earlier (Figure 2h). This allows us to see how learning one polynomial affects knowledge about other polynomials. At the end of epoch zero, we found that all three models were successful at reducing the error on the trained polynomial. At the end of epoch one, again, all models were successful at reducing error for polynomial one, but the error for polynomial zero 1) increased significantly for the regular neural network, 2) only slightly increased for LCNet; and 3) did not increase for feedforward SEM. The results were similar for later polynomials. In general, learning more polynomials caused 1) a severe level of forgetting for the standard neural network, 2) a slight amount of forgetting for LCNet, and 3) no forgetting at all for the SEM model with fully separated representations. A priori, it is not entirely obvious that using random vectors would be sufficient for reducing interference – the projected patterns of these random vectors to the hidden layer might not be orthogonal, as the task-specific parameters (in blue; Figure 2b) were learnable. However, our results empirically showed that indexing different polynomials using random vectors is still effective at reducing interference.

To understand the extent to which prior learning speeds up new learning, we looked at the learning curves while these models went through each individual training observation within every epoch (Figure 2i; Note that the learning rates across models were the same to ensure learning curves are comparable.). For the SEM model with fully separated representations, since there is no re-use of prior knowledge, the learning curve for each polynomial was qualitatively similar. However, for LCNet and the standard neural network, having learned the first polynomial, learning became much faster since they can re-use the previously learned information about the shared term to facilitate new learning.

Overall, there is a tradeoff between representing the shared structure versus eliminating interference across tasks. By using completely separated representations for different LCs, SEM can completely eliminate interference across LCs, but it cannot extract the shared structure. On the other hand, a regular neural network can learn the shared structure, as it uses overlapping representations for all tasks, but it suffers from catastrophic interference. Notably, LCNet can store the shared structure in a common neural network using overlapping representations and store the task-specific structure in its LCI module. As a result, LCNet combines the advantages of both representational schemes – it can extract the shared structure across LCs and use it to make new learning more data efficient while overcoming catastrophic interference. This result resonates with findings from the study of cognitive control, showing that separated representations can support parallel execution of multiple tasks (multitasking) at the cost of learning efficiency, while representation of the structure shared across tasks facilitates learning at the cost of inability to multitask due to inter-task interference[31–33]. In practice, an agent needs to balance the risk of catastrophic interference and the amount of shared structure across LCs. These two factors depend on the similarity structure across tasks[34,35], as well as the number of tasks in the environment.

## Simulation 2: Modeling the curriculum effect on schema learning

In this simulation, we demonstrate that LCNet can explain recent findings on curriculum effects on schema learning that depend on having separated representations for different contexts. The effect of interleaved versus blocked curriculum has been extensively studied in the domain of learning and memory [36–44]. Selecting between an interleaved and blocked curriculum determines the autocorrelation structure of the context: In the blocked setting, the ongoing context rarely changes, whereas in the interleaved condition, the context changes very frequently.

In a recent study[9], participants learned a context-dependent state prediction task using either an interleaved or a blocked curriculum. Each state was a small part of a narrative, and a sequence of states formed a coherent event. During training and testing, the participants were prompted to make a two-alternative-forced-choice about the upcoming state. The state transition graph is shown in Figure 3a. At the beginning of each trial, participants observed if the event was in a cafe context or a bar context. Unbeknownst to the participants, this location (cafe vs. bar) was the context-indicative signal (CIS) that determined the state-transition structure for this trial. Figure 3b shows an example sequence from the graph, conditioned on the cafe context. Overall, the participants' goal was to learn how events tend to unfold in the two contexts.

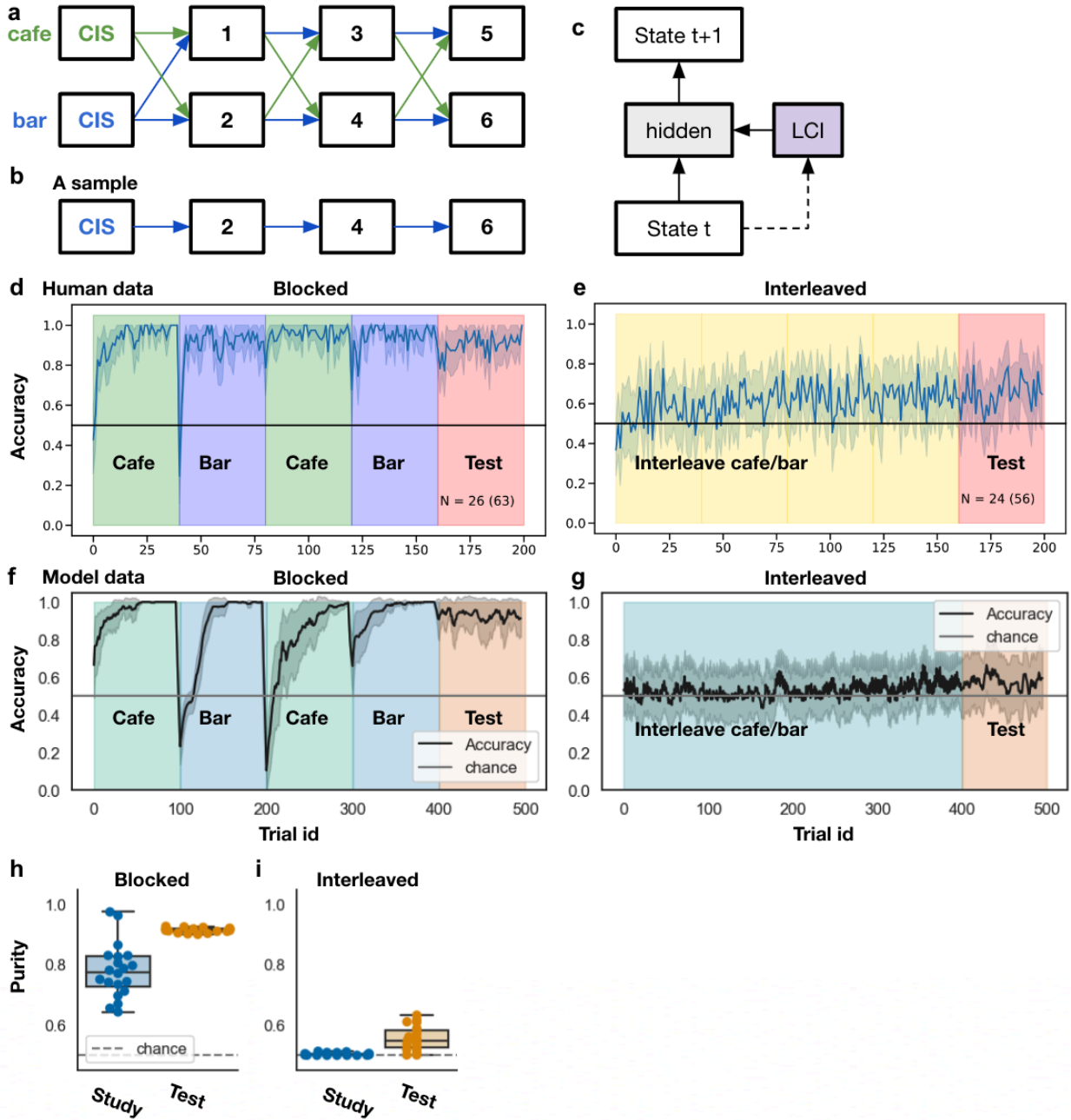

Figure 3. The experiment design and empirical results from ref. 9 and the corresponding simulation results.

a) The state-transition graph (event schemas) used in the experiment. At the beginning of every trial, the participants observe if the event will happen in a cafe or a bar. Unbeknownst to the participants, this (cafe vs. bar) is the context-indicative signal (CIS) that signals the true underlying context which determines the transition structure – in a cafe, the blue transition structure is used, so sequences 1-4-5 and 2-3-6 can occur; in a bar, the green transition structure is used, so sequences 1-3-5 and 2-4-6 can occur. Note that the transition structure is the only difference across the two contexts. The states are aliased in the sense that state *i* in the blue graph is identical to state *i* in the green graph for all *i*. "CIS" = context-indicative signal. b) A sample sequence from the graph conditioned on the "cafe" context.

c) The model architecture. The LCI (latent cause inference) mechanism uses the current state to infer the ongoing LC (see methods details). This is identical to the architecture shown in Figure 4c without the recurrent connection.

d and e) Human performance in the blocked versus interleaved condition. The background colors indicate the training curriculum – blue/green indicates that participants were trained on context 1/2 (i.e., bar/cafe) only. Yellow indicates the participants were trained in the interleaved condition, in which the two contexts alternated from trial to trial. Red indicates the test phase, in which contexts were sampled randomly from trial to trial. During the test phase, participants in the blocked setting performed much better compared to the interleaved condition. This cannot be explained by a standard neural network model since a blocked curriculum would trigger catastrophic interference. Figures from ref. 9.

f) and g) show the model performance over trials in the blocked vs. interleaved condition. Error bands indicate 3SE. N = 20 models per condition.

h) and i) show LCI cluster purity during the test phase in the blocked vs. interleaved condition.

---

The study found that, during the test phase, participants in the blocked curriculum performed much better than participants in the interleaved curriculum (Figure 3d, e)[9]. Notably, a standard recurrent neural network makes the opposite prediction – it performs poorly in the blocked curriculum due to catastrophic interference and performs much better in the interleaved curriculum[9]. Finally, a simplified version of SEM was able to capture the data – this SEM model maintains a transition matrix for every unique LC, which keeps track of the counts of the observed transitions. Then, SEM predicts upcoming states according to the transition counts of the currently inferred LC. When the model hyperparameters were set to promote autocorrelation of LCs in time, LCI was much more accurate in the blocked condition compared to the interleaved condition[9].

Figure 3c shows the architecture of the model used in this simulation. At a high level, the learning objective of the model is to predict "what would happen next." More concretely, at time *t*, once the model gets the current state $x_t$, the LCI mechanism chooses the LC with the maximum posterior probability (see the method section on the Details of the latent cause inference procedure). Then, the model predicts the next state $x_{t+1}$ given $x_t$, and the context vector that corresponds to the selected LC.

We found that LCNet can account for human data. Its performance during the test phase was much better in the blocked condition compared to the interleaved condition (Figure 3f, g). This is because the model's prior on LC inference was set to be sticky; this assumption that contexts are autocorrelated in time has been incorporated in other models of continual learning [3,43,45], including models of this particular task[9,46]. In our case, the stickiness prior matches the autocorrelation structure in the blocked curriculum, where LCs are persistent, compared to the interleaved curriculum, where LCs have low autocorrelation. As a consequence, LCI was much more accurate in the blocked condition (Figure 3h, i). Supplement 2 shows the distribution of the number of LCs inferred in the interleaved condition.

In general, different factors can affect the relative benefits of blocked versus interleaved learning. For example, prior work has shown that – in category learning tasks – the two kinds of curricula can differentially direct people's attention to different dimensions of the stimuli [44]. Providing a comprehensive account of the many nuanced ways in which curriculum manipulations can affect learning is out of scope for this paper. However, in Supplement 3, we further demonstrate that LCNet does not always perform better in the blocked curriculum – instead, its performance depends on how well its stickiness prior matches with the level of autocorrelation of the task environment. As the stickiness decreases, the model's performance during the test phase 1) increases in the interleaved condition and 2) decreases in the blocked condition.

Overall, in this simulation, we found that LCNet can explain human data[9] on how curriculum affects schema learning. Our model qualitatively captured the finding that people performed better in the blocked condition than in the interleaved condition, in contrast to a standard recurrent neural network that made the opposite prediction[9]. Critically, though explaining the human data might seem to depend on separated representations, as the transition structures for the two contexts are highly dissimilar, having a common network did not prevent LCNet from explaining this data.

One limitation of this simulation is that LCI was performed all the time (for all time points and all trials), which is not plausible. In Supplement 4, we present a proof-of-concept simulation in which LCNet is augmented with an episodic memory mechanism that can retrieve previously used LCs in response to sensory cues; the use of episodic memory allows the model to rely less often on full Bayesian inference, while still qualitatively capturing human data.

## Simulation 3: modeling human behavior on naturalistic stimuli

Here, we demonstrate that LCNet can be applied at scale to account for human data and learn ground truth structure from naturalistic videos of daily events, where 1) the structure of the LCs is more realistic and 2) the context-indicative signal (CIS) is implicitly embedded in the sensory input rather than explicitly presented. We used the META (Multi-Angle Extended Three-Dimensional Activities) corpus, a naturalistic dataset generated in a controlled manner[19]. META consists of 25 hours of videos of natural events with a realistic hierarchical structure

(Figure 4a). As its generative structure is known, there are moment-by-moment event labels, which can be thought of as the underlying LCs.

Event segmentation data were collected at two different timescales; while participants were viewing the META videos, they were instructed to press a button to indicate event boundaries for "the largest meaningful units of activity" (i.e., coarse-grain segmentation), or "the smallest meaningful units of activity" (i.e., fine-grain segmentation) [19]. Consistent with prior findings, participants tend to have high agreement on where event boundaries are[10,47–49]. These event boundary data can shed light on LCI in humans.

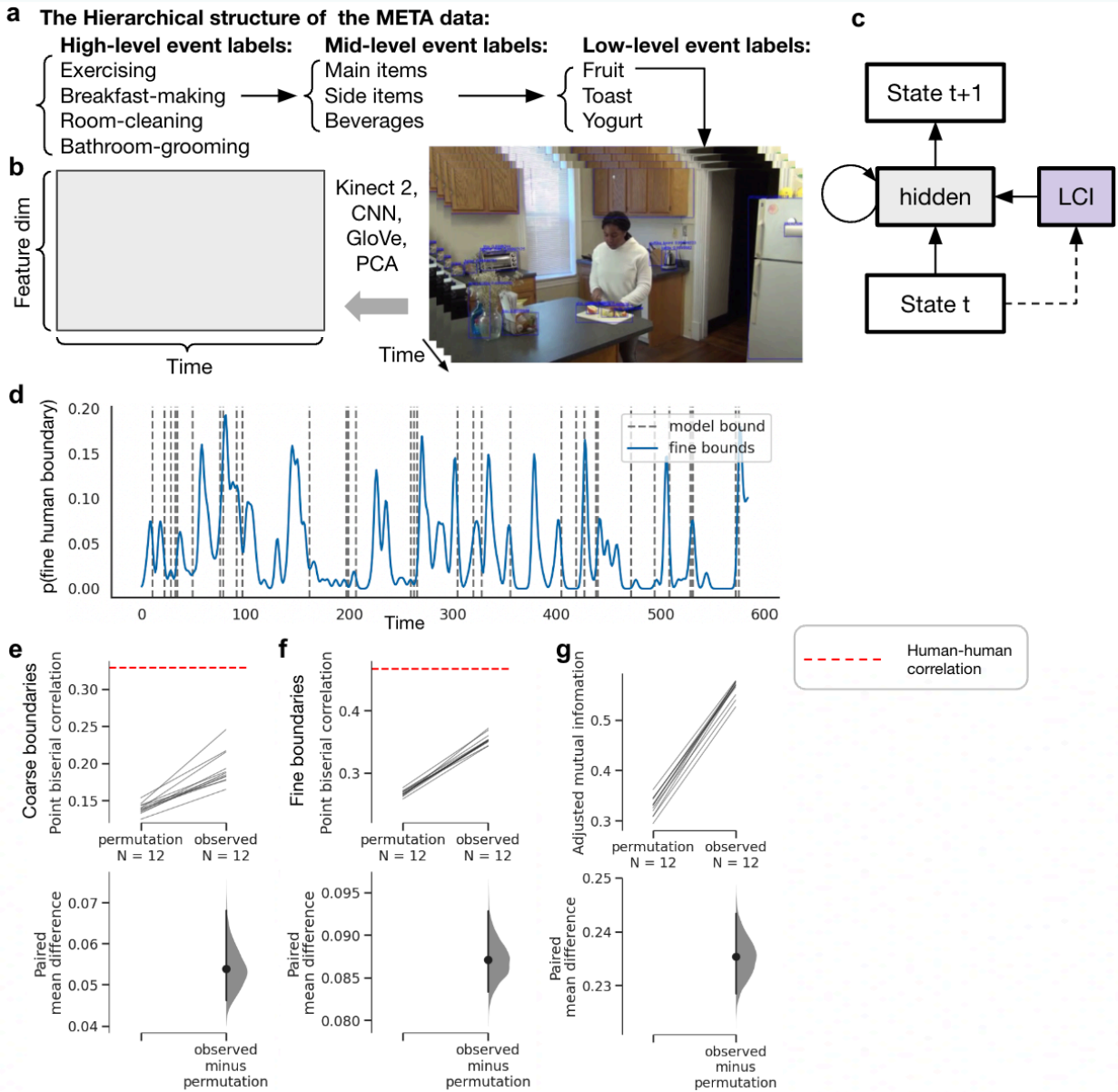

Figure 4. Learning event structure from naturalistic videos

a) The hierarchical structure of the META data set. For the full hierarchical structure, please refer to the META data paper[19].

b) The pre-processing steps used to compress each video frame to a 30D vector. The details of this preprocessing procedure are documented in the META data paper[19].

c) The architecture of a recurrent LCNet. It is identical to the architecture shown in Figure 3c with recurrent connection.

d) The inferred event boundaries versus human boundaries over time for one example video.

e) and f) The correlation between model boundaries and (e) coarse-level / (f) fine-level human boundaries relative to permutation distributions. The red dashed lines mark the mean correlation across all human subjects in the META dataset.

g) The adjusted mutual information over time between inferred LCs versus the ground truth. Panels e, f, and g were generated using dabest package [50]; $N = 12$ models.

---

For feasibility reasons, META video frames were compressed to 30D vectors using a combination of human pose annotation with Microsoft Kinect 2, convolutional neural networks, word embedding models[51], and principal component analysis (Figure 4b)[19]. Conceptually, the compressed representation encodes the body motion of the actors, information about objects in the scene, whether a new object appeared or if some object disappeared, and the correlation of pixel luminance between consecutive video frames. This dimensionality reduction captured 76% of the total variance[19].

In META (and in naturalistic events in general), shared structure exists at multiple timescales. However, the standard version of SEM cannot extract this shared structure as it uses fully separated representations for different LCs. To address this issue, a recent work presented the "SEM 2.0" model, which augments the original SEM by training an additional "generic neural network" on all past experiences in parallel[18]. SEM 2.0 still uses separate networks to handle experiences from different LCs, but when it infers a new LC, it initializes that new neural network using the weights of the generic neural network[18]. This initialization scheme can be viewed as an approximation of the Bayes-optimal initialization for a hierarchical model, since the generic network can be thought of as an average of all past experiences. More details of SEM 2.0 are documented in the methods section for Simulation 3 below.

In our simulation, we used a recurrent LCNet (Figure 4c) with Gated Recurrent Units (GRU; see the Methods section for Simulation 3 for more details)[52]. We used a recurrent neural network (RNN) in this simulation since this is a time series prediction task, and using an RNN to integrate information about recent history in this setting is more cognitively plausible than using a feedforward network[53]. We used a GRU architecture to make the results more comparable to the SEM results, as the GRU architecture was also used in the SEM model[3,18]. The training procedure is the same as SEM 2.0[18] – the objective function is to predict the upcoming state (see method section for Simulation 3 for more detail). At time $t$, the model receives the current scene and predicts the scene that will appear at time $t + 1$. Importantly, there is no feedback on LCI. In Supplement 5, Figure S6 shows the number of LCs inferred over time, which qualitatively matches SEM 2.0, and Figure S7 shows that inferred LCs were indeed reused throughout learning.

To model human data on event segmentation, we extracted event boundaries from our model (operationalized as switch points between inferred LCs) and compared these boundaries with human data. Figure 4d illustrates an example comparison for a single video. The point biserial correlation between the model boundaries and human boundaries (see the methods section for more details) was significant ($p < 0.00001$) (Figure 4e, f), even though there is no supervision on event segmentation and the model objective is to predict the upcoming state. This can be explained in terms of the idea that prediction performance will be most accurate when there is a strong correspondence between inferred LCs and the ground truth event structure; thus, models trained to optimize prediction accuracy will naturally learn to be sensitive to this event structure. To further investigate this, we measured the adjusted mutual information between the inferred LCs and the ground truth event labels over time (see the methods section for more detail) and found a significant correspondence relative to a permuted baseline ($p < 0.00001$) (Figure 4g). This indicates that the inferred LCs captured the ground truth structure of the META videos. That said, the match between LCNet versus human data and ground truth event labels was slightly lower than SEM 2.0 (as SEM 2.0 has several extra mechanisms that we did not incorporate in LCNet). In Supplement 6, we present a quantitative comparison between SEM 2.0. Moreover, as LCNet uses a single neural network shared across all LCs, one can compare the representations across LCs. In Supplement 7, we performed representational similarity analysis (RSA)[54,55] to compare the representational dissimilarity matrix (RDM) computed from LCNet hidden representation of LCNet versus the perceptual similarity RDM computed from META scene vectors and found that they are significantly correlated. Overall, these results show that, like SEM 2.0, LCNet can learn the ground truth event structure and explain human event segmentation data using a predictive objective.

## Discussion

We presented a neural network model that uses a latent cause inference (LCI) mechanism to support learning and generalization in a context-dependent manner. LCNet indexes LCs using vector representations that are approximately orthogonal (see Supplement 1), which is effective at reducing catastrophic interference (Simulation 1). Additionally, by using a single common network that employs overlapping representations for all tasks, LCNet can also extract shared structure across tasks and leverage such structure to learn new tasks more rapidly (Simulation 1). Though SEM can also overcome catastrophic interference with fully separated representations, it cannot optimally reuse prior knowledge when learning a new related task (Simulation 1). Importantly, we showed that the ability to represent shared structure did not hinder LCNet from explaining two sets of findings that SEM explained using fully separated representations across tasks. First, LCNet can account for human data showing that blocked learning curricula can facilitate schema learning (Simulation 2). Standard neural networks incorrectly predict that blocked learning will lead to catastrophic interference[9]. On the other hand, LCNet could explain human data because it does not suffer from catastrophic interference as much, and our experiments show that having a sCRP prior that matches the level of auto-correlation of the environment was important for explaining human learning. Finally, in the META video dataset[19], we found that the inferred LCs fit with the ground truth event labels, and the inferred event boundaries were significantly correlated with human boundaries, even though the model was only trained on predicting the upcoming video frame (Simulation 3). Simulation 3 also allowed us to verify that our modeling framework can be applied to naturalistic data. Overall, these results show that LCNet can reconcile the problem of representing context-specific structure and shared structure for event perception.

While indexing different LCs using random vectors allowed LCNet to avoid catastrophic interference, evidence from human neuroimaging suggests that tasks are sometimes represented in a structured (i.e., non-random) fashion in the brain. For example, one study found that tasks composed of overlapping rules evoke similar patterns of functional connectivity in the fronto-parietal network[56]. To some extent, LCNet can compensate for the lack of structure in its context representations by using the hidden layer to represent shared structure across tasks and learn task-specific information using the weights from the LCI module to the hidden layer. For example, in Simulation 1, the model was able to represent the shared term of the polynomials in the hidden layer. The ability to represent task structure is crucial for explaining human performance: Shared structure across tasks is ubiquitous in naturalistic settings, and studies have shown that humans can extract structure shared across tasks and apply it to novel domains[57,58]. It is unclear if the model's use of orthogonal task representations will prevent it from fully capturing this important aspect of human performance. In the future directions section below, we mention some approaches that would allow subsequent versions of the model to represent shared structure directly in the context layer (in addition to representing it in the hidden layer), which would potentially boost the model's alignment to both neural and behavioral data.

## Related works on context-dependent learning

Several computational modeling papers have addressed context-dependent learning in humans. Models based on standard neural networks without explicit context representation have been used to simulate the development of representations for community structure[59], event structure[60], and people's ability to segment events[53]; in these works, the models were variants of standard feedforward or recurrent neural network models, and the task environments were interleaved. For example, the model in Reynolds et al. (2007)[53] is similar to a standard long short-term memory network (LSTM)[61], where the event representation is similar to the LSTM cell state and the instantaneous output loss is used to gate the cell state. However, standard feedforward and recurrent neural networks are known to exhibit catastrophic interference in blocked settings, where experiences are highly auto-correlated[62–64], so they cannot explain the human data modeled in Simulation 2. One solution to overcome catastrophic interference is to feed explicit context-indicative signals to the network or equip the network with a symbolic context layer[43,45,65]. However, in these models, task labels are provided all the time, so the models do not need to infer the task identity and or detect switches between blocks (i.e., event boundaries). Additionally, the number of tasks in the environment is also given, so that the model can preallocate appropriate neural resources, such as the dimension of the context representation, to represent different contexts. Both of these assumptions are generally not fully valid in real-world settings. The SEM framework use fully separated representations for different contexts to overcome catastrophic interference and use a nonparametric LCI algorithm to i) infer the ongoing task identity and ii) allocate neural resources in an adaptive manner. LCNet goes beyond the standard SEM model to simultaneously represent the shared structure and context-specific structure using a shared network and context vectors.

From the perspective of machine learning, the setting of interest (Figure 1) can be viewed as a continual learning problem[66–69] where agents have to learn a set of tasks sequentially without forgetting previous tasks. In the case of event cognition, both the task identity for the ongoing observation and the number of tasks are unknown. Additionally, observations in natural events are temporally autocorrelated[70]. Effective approaches for continual learning have been an active direction of investigation in cognitive neuroscience[3,42,43,45,65,71–74] and machine learning[67,71,73,75–82]. SEM is similar to a mixture of experts architecture[8], where the LCI mechanism was used to select an appropriate network. If LCI is sufficiently accurate, then a continual learning of $K$ tasks can be decomposed into $K$ single-task learning problems. By contrast, LCNet uses continuous vectors to influence the hidden states, which is similar to continual learning algorithms based on context-dependent modulation [69,83,84]. Though using continuous context vectors to modulate a single shared network might not be as expressive as SEM, LCNet can better leverage from prior knowledge. That said, in this work, the primary goal is to understand and capture empirical data on human learning, rather than seeking an algorithm that performs more effectively on continual learning benchmarks from the machine learning literature.

## Future directions

In the standard version of SEM, LCI has to be performed all the time, which is computationally costly. One way to reduce unnecessary inferences is gating – results from variants of SEM 2.0 suggest that it can be effective to perform full inference only when uncertainty or prediction error is high[18]. We also experimented with a complementary approach using episodic memory. In Supplement 4, we augmented LCNet with a simple episodic memory module[85,86], which allows the model to retrieve a previously inferred LC given the current observation, instead of performing full inference; using the task presented in Simulation 2 as a testbed, we were able to demonstrate that our model can work much more efficiently without compromising its ability to account for human data. Additionally, a recent computational model suggests that episodic memories can play a critical role in shaping the ongoing task representation[46,87]. Overall, combining models of event segmentation, such as LCNet and SEM, with episodic memory would also allow us to address data on how event structure influences memory encoding[10,47,48,88] and retrieval[89].

An important limitation of SEM-based models, including LCNet, is that inferred LCs are represented in an unstructured list. Concretely, SEM represents the inferred LCs as a list of networks, and LCNet represents LCs as a list of random vectors. The computational cost of full LCI (evaluating the posterior over LCs) is linear with respect to the number of inferred LCs, which is expensive. Additionally, there is no way to gracefully "merge" context representations that turn out to be redundant after they have been split apart into different LCs – they continue to exist in structurally distinct form for all subsequent time points. One potential way to comprehensively address these issues is to learn a feature space of LC representations where distances reflect the relations across tasks, instead of representing all LCs with distinct random vectors. This idea dates back to classic parallel distributed processing models of semantic cognition, where activation space gradient descent was used to find hidden representations of objects[90]. Recent results have demonstrated that this algorithm can identify context representations to support controlled semantic processing[91] and context-dependent learning[87,92]. Importantly, organizing LC representations in a structured feature space would potentially enable graceful similarity-based generalization.

Lastly, an important property of real-world settings is that multiple LCs can be active at the same time to affect the ongoing events[93]. For example, suppose the agent previously observed some coffee-drinking events and some helicopter-riding events. Then, if the agent encounters a person who is drinking coffee on a helicopter, one reasonable approach is to activate the two LCs used previously to process coffee-drinking events and helicopter-riding events.

Recently, large language models using a transformer-based architecture achieved impressive results in many application domains[94,95]. Though these models are powerful, they do not always account for successes, and in particular, failures of human cognition[96]. This is potentially because their computational mechanisms are systematically different from human brains[96]. In the future, it would be informative to understand the potential contribution of an explicit LCI module to transformer-based architectures.

In conclusion, LCNet addresses a key limitation of the SEM framework regarding how to balance the ability to represent multiple LCs while managing interference across them. We view this as a useful step towards a more neurally plausible model of LCI in the context of event cognition.

# Methods

## Latent cause representation in SEM versus LCNet

The most important distinction between SEM and LCNet is the format in which latent causes are represented. In SEM, each latent cause is represented by a unique neural network. At time t, if the current observation triggers a new LC, a new neural network will be initialized and used to process the current observation. Otherwise, if the current observation activates an existing LC, the corresponding network will be used to process that observation. This algorithm can be written as follows:

Algorithm 1

**Processing steps, SEM**
Given an observation at time $t$, $x_t$, assuming the model has $K$ latent causes so far
If the model infers a new latent cause:
    initialize the ($K$+1)-th network using random weights
    use the ($K$+1)-th network to process $x_t$
else, the model assigns $x_t$ to the $k$-th network ($k$ in [0, $K$]):
    use the $k$-th network to process $x_t$

By contrast, LCNet represents LCs with random vectors; there is a single network that is shared for processing observations regardless of the LC assignment, which enables that network to extract shared structure in the environment. Concretely, if the current observation triggers a new LC, then LCNet samples a new random vector, which will be fed to the network. Otherwise, if the current observation activates an existing LC, the corresponding random vector will be fed to the network. This algorithm can be written as the following (the differences between Algorithm 1 and Algorithm 2 are marked in blue):

Algorithm 2

**Processing steps, LCNet**
Given an observation at time $t$, $x_t$, assuming the model has $K$ latent causes so far
If the model infers a new latent cause:
    sample a random vector $c_{K+1}$, and feed it to the hidden layer
else, the model assigns $x_t$ to the k-th network ($k$ in [0, $K$]):
    feed $c_k$ to the hidden layer
Use the whole network to process $x_t$

Details of the latent cause inference procedure

In both SEM and LCNet, LCI determines i) whether a known LC best explains the current observation or ii) if this observation has to be assigned to a new LC. The inference procedure combines the likelihood and prior to obtain a posterior distribution over all known LCs. Then, the LC with the maximal posterior is selected.

LCNet computes the likelihood by comparing all of the known LCs in terms of how well they explain the current observation. Concretely, assume that K LCs have been generated so far, and $c_1, c_2, ... c_K$ are the corresponding context vectors. Let f be an instance of LCNet, which maps an observation $x_{t-1}$ and a context vector $c_k$ to the predicted next state $\hat{x}_t$. Then given an observation $x_t$, the model computes the losses over all LCs. The loss of the k-th LC is the following:

$$\text{MSE Loss}(t, c_k) = \text{MSE Loss}(f(x_{t-1}, c_k), x_t)$$

We compute the loss of a new LC using a fixed preallocated random context vector. Then we take the softmax of the negative loss over LCs and treat it as the likelihood, which means the maximal likelihood LC will be the LC with the lowest loss. Note that, at time *t*, the goal of the model is to predict $x_{t+1}$ – but since the model hasn't observed $x_{t+1}$ yet, the likelihood is computed by using $x_{t-1}$ to predict $x_t$ (rather than using $x_t$ to predict $x_{t+1}$).

Similar to SEM, we use the sticky Chinese Restaurant Process (sCRP) prior[3,97,98]:

$$P(z_t = k \mid z_{1:t-1}) \propto \begin{cases} \text{count}_k + \lambda \mathbb{I}\left[z_{t-1} = k\right] & \text{if } k \leq K \\ \alpha & \text{if } k = K+1 \end{cases}$$

where $z_t$ is the inferred LC at time *t*, *k* is the index for the LCs, *K* is the number of distinct LCs inferred so far, I is the indicator function, $\text{count}_k$ is the number of observations assigned to the *k*-th LC previously, alpha is the concentration parameter that controls the chance of inferring a new LC index by *K*+1, and lambda is the stickiness parameter that boosts the previously used LC. Under the sCRP prior, LCs that were used to explain more observations in the past are boosted, and the currently active LC is also boosted.

At time *t*, the posterior distribution over LCs is computed by combining the sCRP prior with the likelihood. Similar to SEM, we use the local maximum a posteriori (MAP) approximation[98,99] – the LC with the maximal posterior probability is selected to generate the prediction of the upcoming state.

Note that the LCI framework assumes that event segmentation is triggered by switches of the inferred LCs (Figure 1). This view is an extension of a classic view, assuming that event segmentation is triggered by high sensory prediction error[53]. Though sensory prediction error can cause switches of inferred LCs, the switches of inferred LCs can be caused by other factors, such as uncertainty[100,101], which can be dissociated from sensory prediction error. Recent modeling evidence also suggests that allowing uncertainty to drive LCI allows the SEM framework to better explain human event segmentation data, compared to the sensory prediction error view [18].

## Simulation 2 methods details

### Stimulus representation

Each state in the graph (Figure 3a) is represented using a fixed random vector. At time $t$, the model receives a smoothed average of the current state and the previous state. Namely, the current input = $w$ x current state + (1 - $w$) x previous state. This ensures that the model has access to recent stimulus history, which allows us to make the simplification of using feedforward neural networks. Conceptually, this means that a recency-weighted representation of the stimulus history is available to the model. The context-indicative signals (CIS; cafe versus bar) at the beginning of each trial are regular states, which were also represented as two fixed random vectors. Therefore, the CIS in our simulation is just a regular observation, similar to the human experiment.

### Unpredictable transitions

As in a previous simulation using a simplified version of SEM[9], Bayesian inference was turned off at the transition from the CIS to state one vs. state two, which is inherently unpredictable (see Figure 3a). This implements the claim that people are able to learn which transitions are unpredictable, and that prediction errors are down-weighted for unpredictable transitions when making inferences (as hypothesized by normative models of learning[102,103]).

### Cluster purity to measure the inferred latent causes versus ground truth

To measure the accuracy of the inferred LC, we computed cluster purity of the inferred LCs, which is a metric commonly used for evaluating the quality of a clusterin [104].

$$\text{Purity}(\hat{C}, C) = \frac{1}{n} \sum_{k=1}^{K} \max_{j=1}^{J} |\hat{C}_k \cap C_j|$$

where $\hat{C}$ is the vector of LC assignments, and C is the true LC (context) ID for each observation. n is the total number of observations. $K$ is the number of LCs inferred so far. $J$ is the number of true LCs, which is 2 in this simulation. $\hat{C}_k$ is the set of observations assigned to LC $k$. $C_j$ is the set of observations that belong to true LC $j$.

Conceptually, purity measures the percentage of observations assigned to each inferred LC belonging to the same LC. A purity score of 1 means that all observations in each inferred LC belong to the same true LC, while a score of 0 means that the inferred LC over observations does not match the LCs at all.

## Simulation 3 methods details

Gated Recurrent Unit

In Simulation 3, due to the complexity of time series prediction for naturalistic video, we used a widely used recurrent neural network (RNN) architecture known as the Gated Recurrent Unit (GRU)[52], which is also used in the SEM model[3]. The temporal dynamics of a GRU-based RNN are governed by the following equations.

$$z_t = \sigma(W_z \cdot [h_{t-1}, x_t])$$
$$r_t = \sigma(W_r \cdot [h_{t-1}, x_t])$$
$$\tilde{h}_t = \tanh(W \cdot [r_t \odot h_{t-1}, x_t])$$
$$h_t = (1 - z_t) \odot h_{t-1} + z_t \odot \tilde{h}_t$$

where $x_t$ and $h_{t-1}$ are the current input and the previous hidden state. Square brackets indicate vector concatenation. $W_z$, $W_r$ are the weights mapping from the previous hidden state and current input to the current read gate $r_t$ and the write gate $z_t$. The read gate controls the level of impact of the previous hidden state on the proposed change of the current hidden state $\tilde{h}_t$. Note that in a GRU, the forget gate is simply $1 - z_t$, so that the writing of new information and the forgetting of old information are synchronized. Concretely, the current, updated hidden state $h_t$ is simply an interpolation between $h_{t-1}$ and $\tilde{h}_t$ weighted by the forget gate and the write gate.

Point biserial correlation between the model boundaries versus human boundaries

Point biserial correlation is used to characterize the correlation between the human boundary probability at time *t* for all *t* (a continuous-valued vector) and whether there is a model boundary at time *t* for all *t* (a zero-one valued binary vector, where one indicates that the model inferred a boundary at time *t*):

$$r_{pb} = \frac{M_1 - M_0}{s_n} \sqrt{\frac{n_1 n_0}{n^2}}$$

where $M_1$ is the mean human boundary probability value when the model inferred a boundary, $M_0$ is the mean human boundary probability value when the model did not infer a boundary, $n_1$ is the number of event boundaries inferred, $n_0$ is the number of time points without an inferred event boundary, $n$ is the number of total time points ($n = n_1 + n_0$) and $s_n$ is the standard deviation of human boundary probability values over time.

Similar to the analysis in SEM 2.0[18], since the largest and the smallest achievable values of point biserial correlation depend on the number of boundaries the model inferred (i.e., the number of ones in the zero-one valued vector), we scaled the raw correlation value by the largest and the smallest achievable values.

### Adjusted mutual information between the inferred latent causes versus the ground truth event labels

Compared to real-life events, a unique advantage of META as a controlled video dataset is that we have the underlying event labels used to generate the META videos. These event labels over time can be viewed as the ground truth of the LCs. The finest level of event labels available in META is the 45 subevent classes.

Following the way SEM 2.0 was evaluated, we computed the mutual information between the inferred LCs versus the ground truth event labels[18].

$$\text{AMI}(C, \hat{C}) = \frac{\text{MI}(C, \hat{C}) - E[\text{MI}(C, \hat{C})]}{\max(H(C), H(\hat{C})) - E[\text{MI}(C, \hat{C})]}$$

where $C$ and $\hat{C}$ are the true event labels and inferred LCs, MI is the mutual information function, $H$ is the entropy function, and $E$ denotes the expectation.

## Acknowledgment

This work was supported by a Multi-University Research Initiative grant to K.A.N., J.M.Z., S.J.G., and U.H. (ONR/DoD N00014-17-1-2961). We thank the two anonymous reviewers for their constructive feedback.

## Author contributions

Q.L. contributed to:
Conceptualization, Data curation, Formal analysis, Investigation, Methodology, Software, Validation, Visualization, Writing – original draft, Writing – review & editing

T.T.N. contributed to:
Data curation, Formal analysis, Software

Q.Z. contributed to:
Conceptualization, Writing – review & editing

U.H, T.L.G., J.M.Z., S.J.G., contributed to:
Conceptualization, Funding acquisition, Writing – review & editing

K.A.N. contributed to:
Conceptualization, Funding acquisition, Investigation, Supervision, Writing – review & editing

## Competing interests

The authors declare no competing interests.

## Data availability statement

All data and code are publicly available:
Simulation 1 and Simulation 2: https://github.com/qihongl/LCNet
Simulation 3: https://github.com/qihongl/meta-model
META data[19] used in Simulation 3: https://osf.io/3embr/

# Supplementary information

Supplement 1

**Inter-context correlation as a function of the dimensionality of the random context vector**

We used random vectors to index different latent causes to ensure that the representations for different latent causes would be approximately orthogonal. This idea is based on the fact that high dimensional random vectors, sampled from Gaussian distribution, are close to orthogonal. Here, we measure how orthogonal random vectors are as a function of dimensionality and the number of vectors (i.e., the number of latent causes).

Through a simple simulation, we found that – as the dimension increases – the values of pairwise correlation rapidly concentrate around zero (Figure S1a). Additionally, as the number of random vectors increases, the maximum pairwise correlation across any two vectors gets higher (Figure S1b), though the mean is relatively stable (Figure S1c).

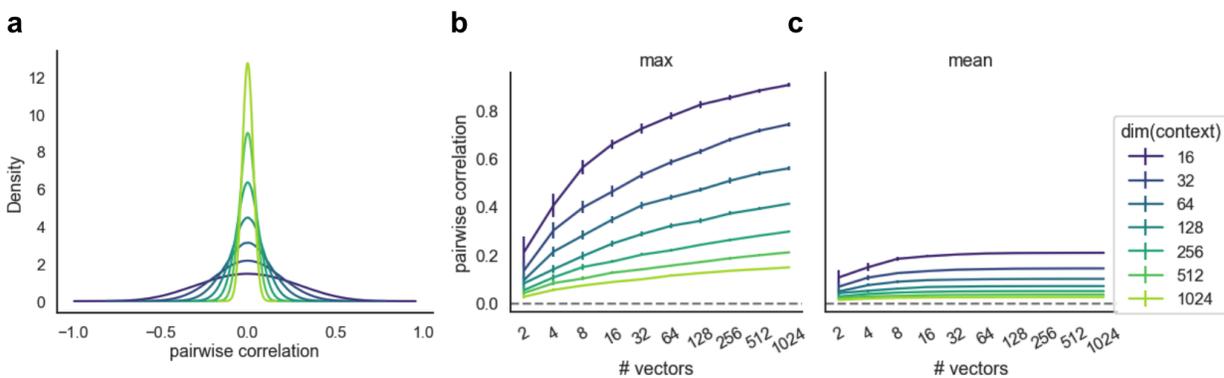

Figure S1

Statistics of pairwise correlation of random vectors as a function of vector dimension and the number of samples/vectors.
a) As the dimension increases, random vectors are more orthogonal.
b) As the number of vectors increases, the maximal pairwise correlation asymptotes. This upper bound is lower in higher dimensional space.
c) As the number of vectors increases, the mean pairwise correlation is relatively stable. Again, in high dimensional space, most vectors are close to being orthogonal. Error bar = 2SE across 50 simulations.

Supplement 2

**In the interleaved condition in Simulation 2, some models "over-split" and some models "under-split."**

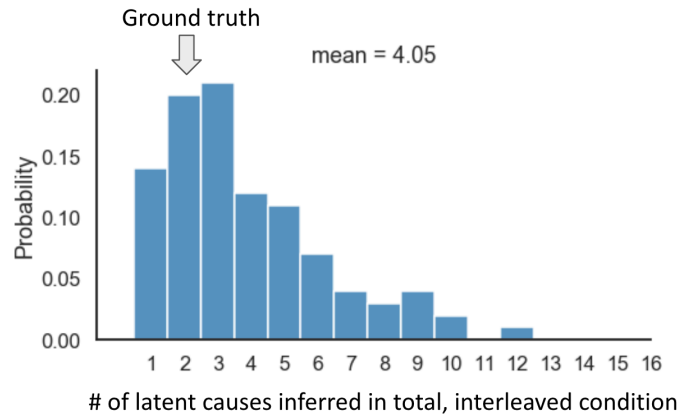

Figure S2

The distribution of the total number of latent causes inferred in the interleaved condition across ($N$ = 100) models. Some models "over-split" (inferred more than two latent causes), and some models "under-split" (inferred only one latent cause). The main issue with models in the interleaved condition is that the correspondence between the inferred latent causes versus the ground truth latent causes was highly inaccurate (Figure 3h and i).

Supplement 3

**The effect of the stickiness parameter on task performance in Simulation 2**

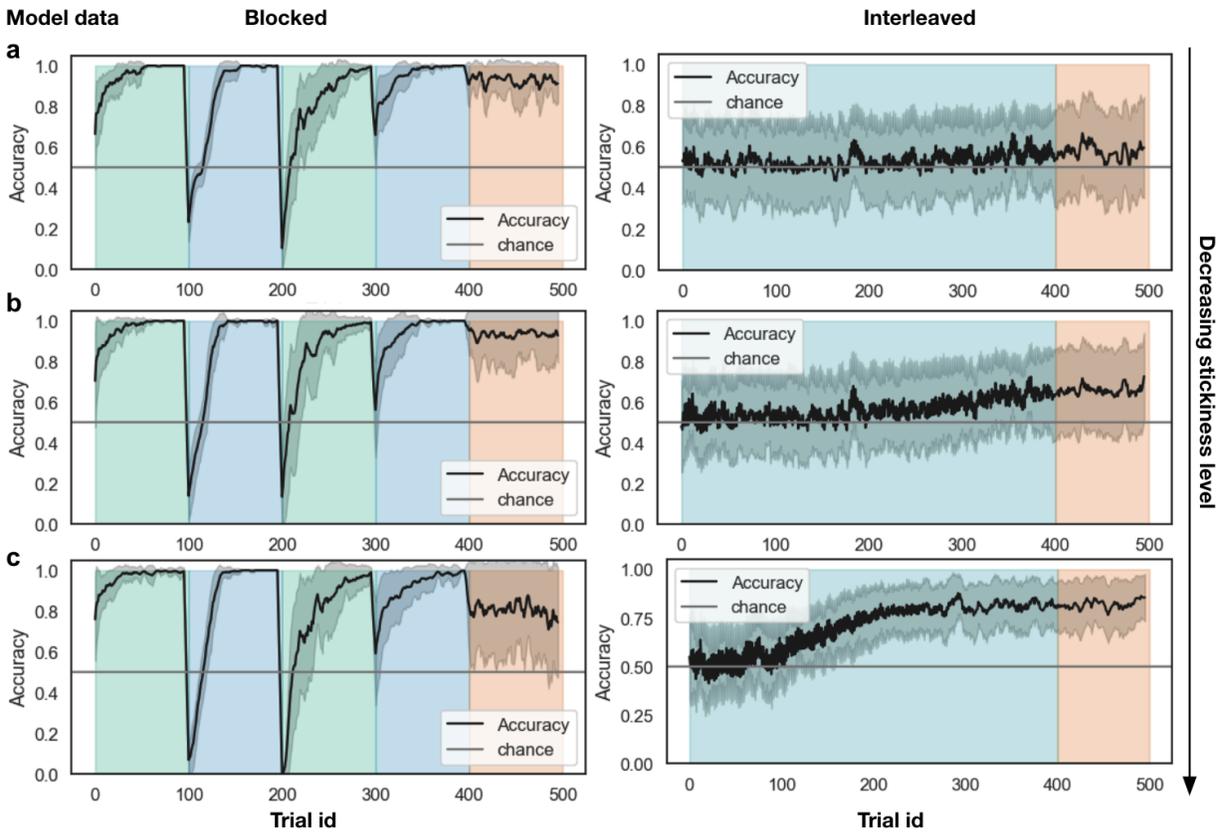

Figure S3

As the stickiness parameter decreases from 32 (a) to 8 (b) and then to 0 (c), the performance of LCNet during the test phase i) increases in the interleaved condition and ii) decreases in the blocked condition. This shows that the relative advantage of blocked over interleaved learning is a function of how well the model's prior on temporal auto-correlation in the environment matches with the ground truth. Note that panel a is the same as Figure 3f and 3g in the main text. Error bands indicate 3SE. $N = 20$ models per condition.

Supplement 4

**Episodic memory as a shortcut to full latent cause inference**

Full LCI in both SEM and the standard LCNet involves evaluating the posterior over all LCs, which is computationally expensive. We hypothesized that humans can leverage episodic memory to economize on LCI, by recalling the LC that was inferred previously for a particular sensory observation. Here, we present a proof-of-concept simulation using the task in Simulation 2 as a testbed.

Concretely, we equipped LCNet with an episodic memory buffer that maps observations to LCs (Figure S4a); this buffer acts as an efficient "shortcut" to the laborious full inference process, similar to the idea of amortization [105]. The episodic memory mechanism stores previously encountered {observation, inferred LC} pairs. This episodic memory "shortcut" is turned on when its predictions become consistent with full inferences; when prediction error is too high, full inference is turned on and episodic memory is turned off. For additional detail, see the next section ("the implementation of episodic memory").

Results from Simulation 2 show that models with episodic memory also qualitatively capture the human behavioral results – during the test phase, prediction performance was much better in the blocked condition compared to the interleaved condition (Figure S4b, c). Again, the reason is that LCI accuracy was much lower in the interleaved condition (Figure S4d, e). More importantly, we found that LCNet with episodic memory can save 96.20% of full inferences (Figure S5a, b) while still capturing the human data. With episodic memory, full inferences selectively happen at the first two block switches. Then episodic memory is sufficient for retrieving the proper LC for the rest of the task.

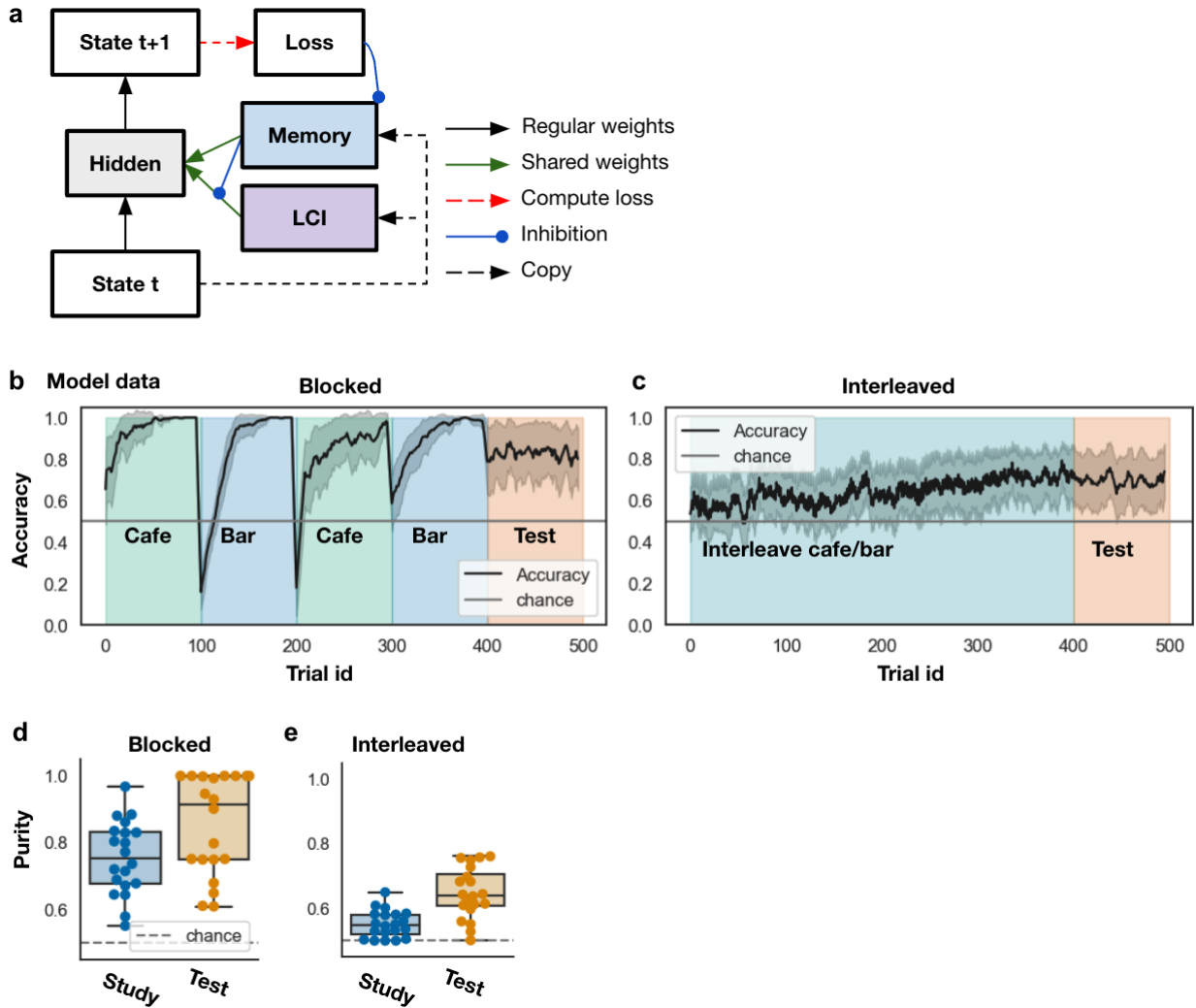

Figure S4

a) The model with the episodic memory mechanism. When the model retrieves a previously inferred latent cause from episodic memory, full LCI is suppressed. When the model uses retrieved LC and the current loss is too high, the model suppresses episodic retrieval and switches back to relying on full latent cause inference. "LCI" = latent cause inference.
b) and c) show the model performance over trials in the blocked vs. interleaved condition.
d) and e) show LCI cluster purity during the test phase in the blocked vs. interleaved condition for the model with episodic memory.

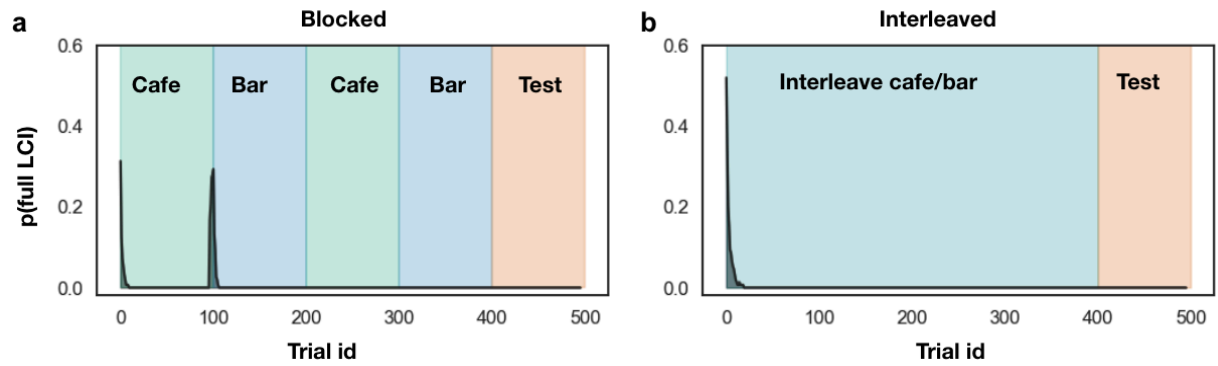

Figure S5

A large amount (96.20%) of full inferences can be saved with an episodic memory buffer. In the blocked condition (a), full inference mainly occurs at the first two event boundaries. In the interleaved condition (b), full inference mainly peaked at the beginning of the experiments. $N = 20$ models per condition

The implementation of episodic memory

Episodic memory is implemented as a key-value dictionary, a simple proxy of content-based associative memory used for a wide range of cognitive tasks[85,86]. In this framework, the key is used for memory search, and the value is used to store the content. To encode a new memory, the model stores the inferred LC (from the full inference procedure) as the value, paired with the current observation as the key. To retrieve a memory, it performs a one-nearest-neighbor (1NN) lookup[85,106] using the current observation as the key.

To ensure that outdated information does not lead to interference[107,108], we implemented a simple form of forgetting – the buffer is a queue that only keeps the most recent $M$ (= 2) memories for each unique observation. M is chosen to be the smallest possible number that does not negatively affect model performance. To incorporate the assumption that the hippocampus has a narrow generalization gradient[63,72,109–113], retrieval was designed to be conservative – given an observation $x$, it will return a LC if and only if there are at least $M$ entries (LCs) associated with x, and they are all the same (Figure S5).

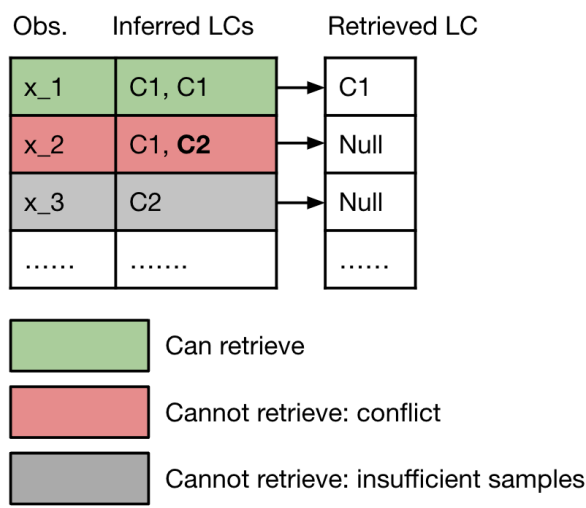

Figure S5
- An episodic memory buffer that keeps the most recent $M$ (= 2) latent causes for each observation $x$. Given $x_1$, it will retrieve C1 because it stored 2 latent causes associated with $x_1$, and all of them are the same. Given $x_2$, it will not retrieve anything because there is a conflict (i.e., C2) across stored memories. Given $x_3$, it will not retrieve anything because it has not stored enough samples for this observation.

Finally, we need to specify when to use episodic memory versus full LCI (Algorithm 3). The rule is simple – given an observation $x$, if an episodic memory is retrieved, the model will use the retrieved LC and suppress the full inference procedure. Once episodic memory is activated for $x$, it will remain active for *x* until the model experiences a high loss for *x*, in which case episodic retrieval for *x* will be turned off, and the row corresponding to *x* in the episodic buffer will be cleared. The model tracks the running mean and standard deviation of the losses conditioned on the observations when full inference is active. The loss is considered to be too high (i.e., loss/prediction error peak) if it is higher than the mean plus three standard deviations, estimated from the history of losses for this observation.

Algorithm 3
**Handoff between episodic memory and full latent cause inference**
**Given** an observation $x_t$
Retrieve episodic memory based on $x_t$
**If** a latent cause, denoted by $C_m$, is retrieved:
    $x_{t+1}$ = model $(x_t, C_m)$
    **If** there is a prediction error (loss) peak:
        clear the memory buffer for *x* (which will turn off episodic memory for *x*)
**else**:
    $C_f$ = full_inference($x_t$)
    memory_buffer.store($x_t$, $C_f$)
    $x_{t+1}$ = model $(x_t, C_f)$
    record the loss and update the mean and standard deviation of the loss

Supplement 5

**Exploratory analysis of inferred latent causes**

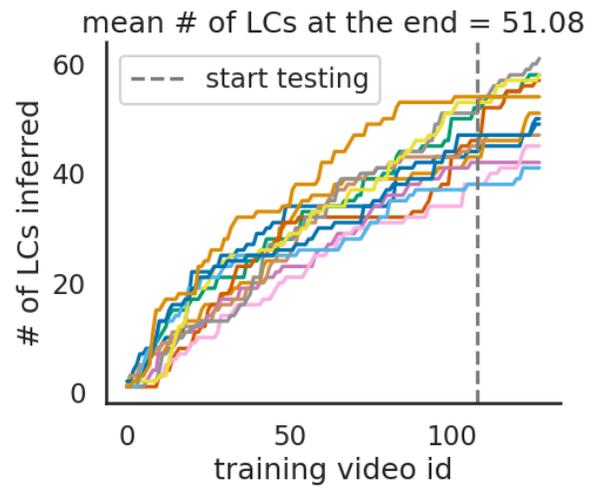

Figure S6

Number of inferred events over time. Each curve is one model. Overall, the number of inferred latent causes grows linearly as the model views more videos.

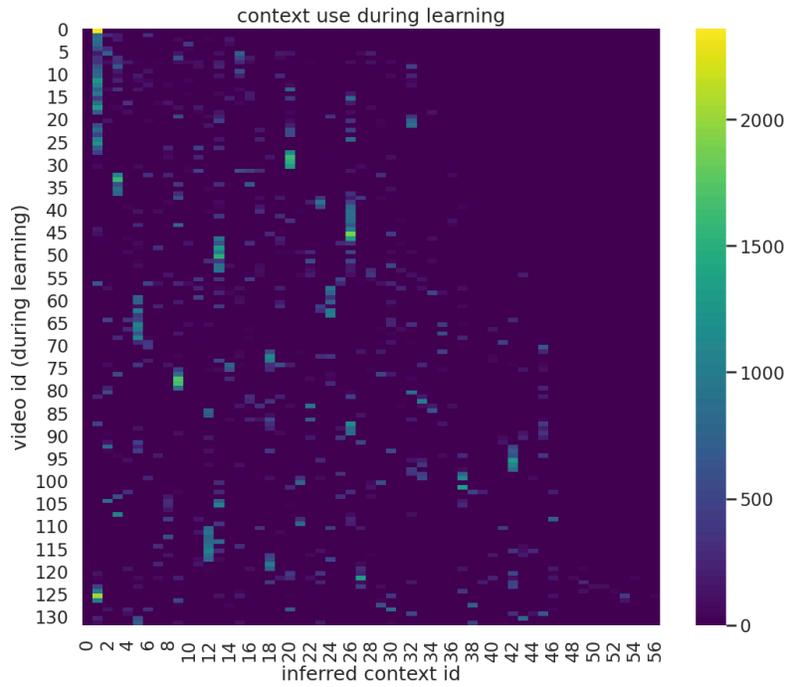

Figure S7

Evidence of reusing latent causes from one example. As LCNet was being trained on META, it reused previously generated latent causes. Here is a heatmap of the number of time points assigned to each of the latent causes for each training video for one example model. A row in this matrix tells us the distribution of latent causes within a video. A column in this matrix tells us how a latent cause was used throughout the training.

Supplement 6

**A comparison between LCNet and SEM 2.0 on Simulation 3**

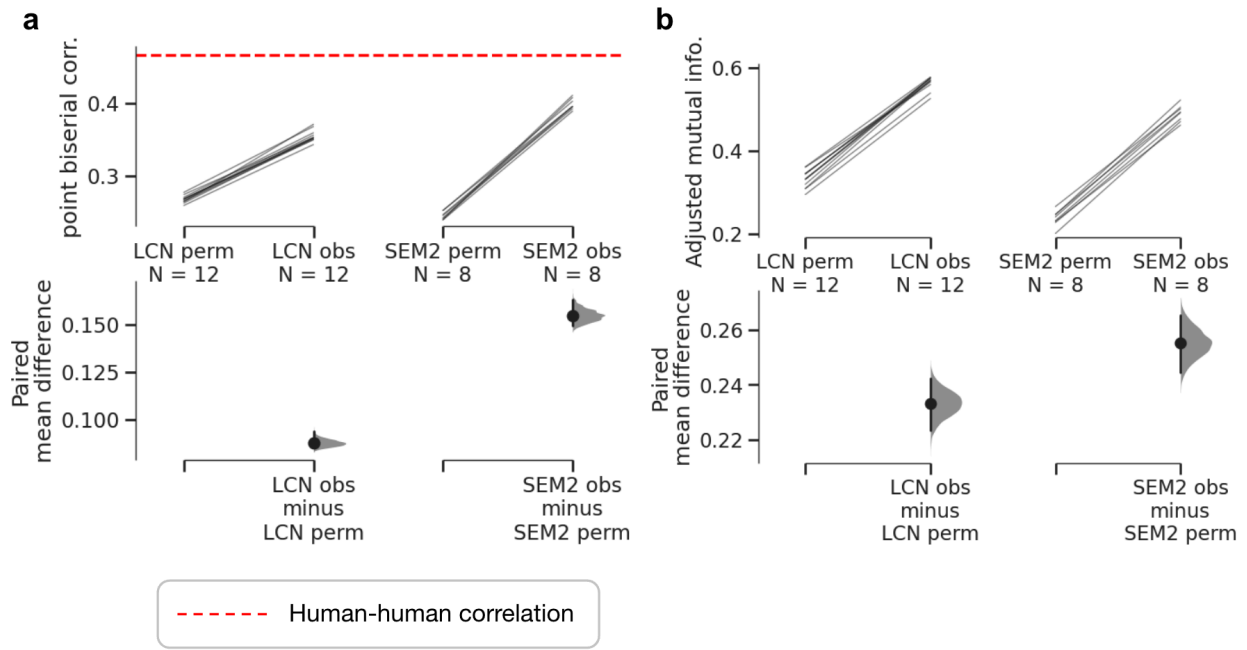

Figure S8

A comparison between LCNet vs. SEM 2.0 shows that LCNet is slightly worse compared to SEM 2.0 in terms of (a) point biserial correlation with human data and (b) adjusted mutual information with the ground truth low-level event labels. This is potentially due to several additional mechanisms in SEM 2.0, such as the fact that it tracks the variance of prediction error in an LC-specific manner, as well as other differences in the implementation details.

Supplement 7

For each low-level event category (i.e., the leaves in the ground truth hierarchy shown in Figure 4a), we compute the mean activity pattern of the recurrent hidden layer. Then we compute the representational dissimilarity matrix (RDM)[54,55] over low-level event categories (Figure 4i). The task RDM is computed using the scene vectors (Figure 4h). The result shows that the correlation between the model RDM and the task RDM is significant (Figure 4j; $p < 1e-5$).

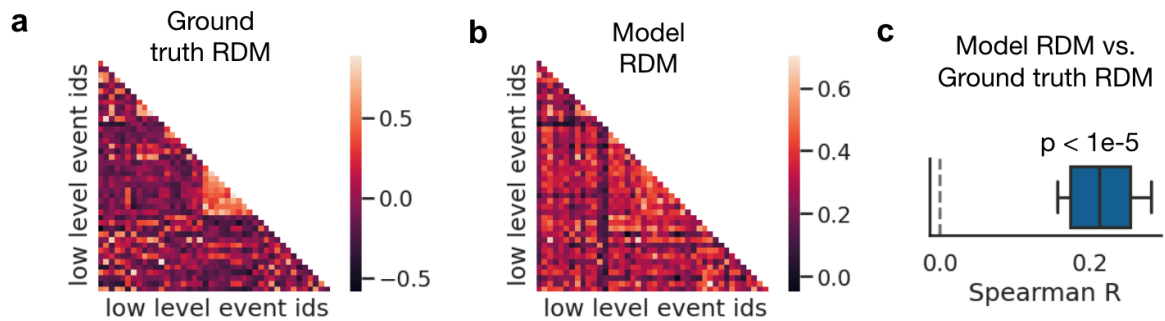

Figure S9

a) The representational dissimilarity matrix (RDM) for the mean patterns averaged within each low-level event.
b) The RDM for the LCNet hidden state averaged within each low-level event.
c) The correlation between the model RDM and the ground truth RDM.

Supplement 8

**Hyperparameters**

We implemented the model in PyTorch[114,115]. In Simulation 1, the number of hidden units and learning rate were equated across models for model comparison. Minor changes in these parameters did not influence the general pattern that i) LCNet suffers less catastrophic interference compared to a regular neural network and ii) LCNet learns new tasks faster than SEM (see Supplement 9).

In Simulation 2, the goal is to show that there is a setting where a blocked curriculum allowed the model to perform better during the test phase, compared to an interleaved curriculum. Therefore, we optimized stickiness and concentration by a hyper-parameter grid search, and used parameters that maximized the model performance in the blocked curriculum. Additionally, we showed that, when using a lower stickiness value, LCNet can perform better in the interleaved condition (Supplement 3). The number of hidden units and learning rate were chosen so that the model did not require too many trials in the experiment to learn the tasks. Minor changes in the number of hidden units and learning rate did not influence the main finding regarding the impact of stickiness on the relative performance between blocked and interleaved conditions.

In Simulation 3, similar to how these parameters were chosen in SEM 2.0[18], the stickiness and concentration parameters were chosen to ensure the number of event boundaries, mean event duration, and average total number of latent causes of the model qualitatively matched with human data. The number of hidden units and the optimizer were also matched with SEM 2.0 to facilitate model comparison. Finally, in all three simulations, we set the dimension of the context vectors to 128 to ensure sufficient orthogonality without adding too many parameters to the model (Supplement 1).

We used the following hyperparameters for the three simulations:
Simulation 1:
- Optimizer: Adam[116]
- Learning rate: $1 \times 10^{-5}$
- Number of hidden units: 128
- The dimension of the context vectors: 128
- The dimension of the context-indicative signal: 128
- Nonlinearity for hidden units: ReLU

Simulation 2:
- Optimizer: Adam[116]
- Learning rate: $4 \times 10^{-3}$
- Number of hidden units: 128

- *w*, the weight on the current state when computing the running average of the input: 0.8
- The dimension of the context vectors: 128
- sCRP prior
    - Stickiness: 32
    - Concentration: 0.5
- Nonlinearity for hidden units: Sigmoidal

Simulation 3:
- Optimizer: Adam[116]
- Learning rate: $1 \times 10^{-3}$
- Number of hidden units: 16
- The dimension of the context vectors: 128
- sUP prior
    - Stickiness: 4
    - Concentration: 1
- Nonlinearity for hidden units: we used the standard nonlinearity in Gated Recurrent Units (GRU) specified in the GRU methods section.

Supplement 9

**Additional manipulation of hyperparameters in Simulation 1**

Here, we conduct additional experiments to ensure that minor changes of the two hyperparameters, the number of hidden units and learning rate, do not influence the main findings that i) LCNet suffers less catastrophic interference compared to a regular neural network and ii) LCNet learns new tasks faster than SEM. In general, decreasing/increasing the number of hidden units or learning rate decreases/increases the learning efficiency for all three models.

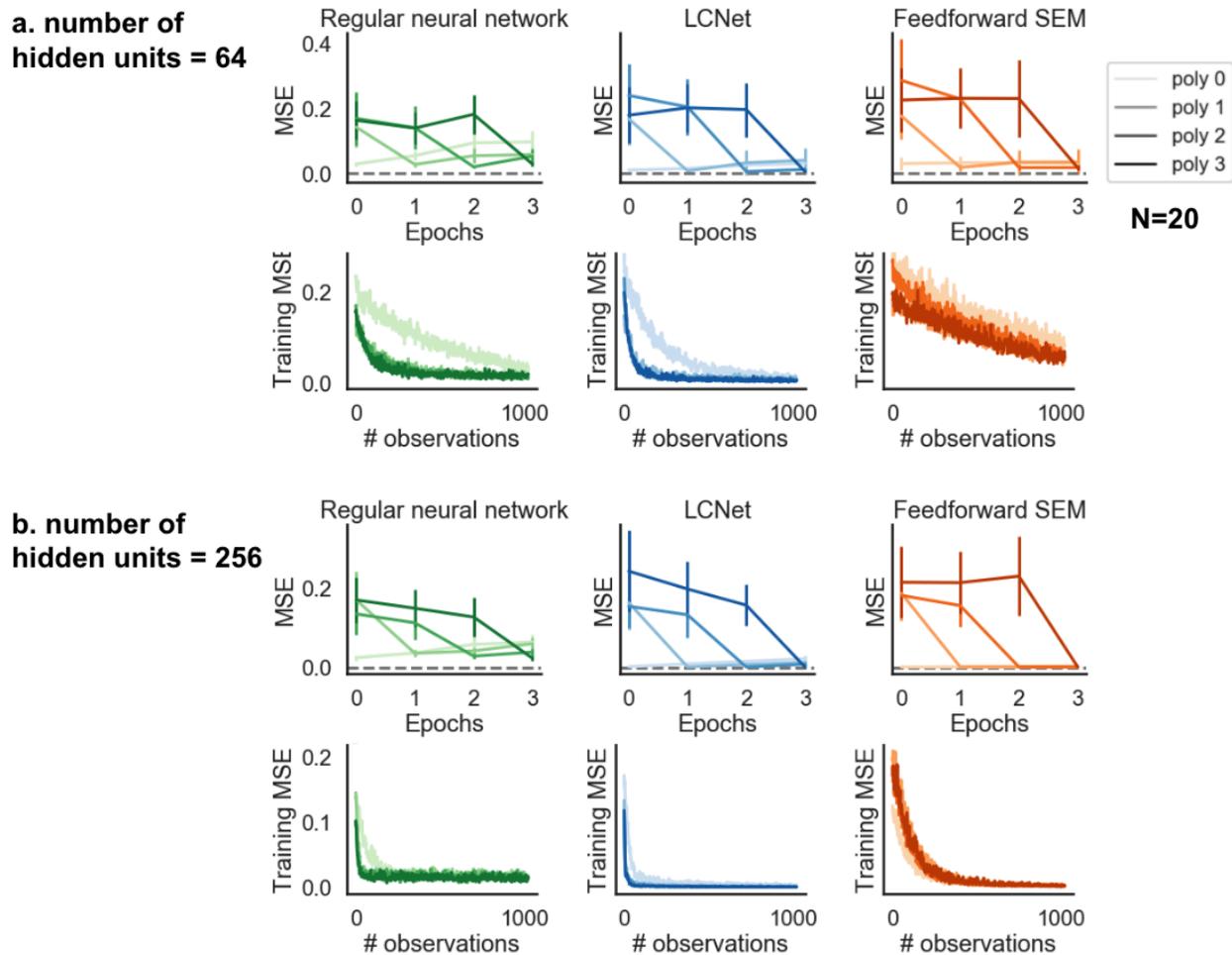

Figure S10

Here, we report the main analyses (see Figure 2h, 2i) when the number of hidden units (which is 128 in the main paper) is changed to 64 (a) and 256 (b). The same pattern reported in the main paper qualitatively holds. Error bars indicate 3SE. $N = 20$ models per condition.

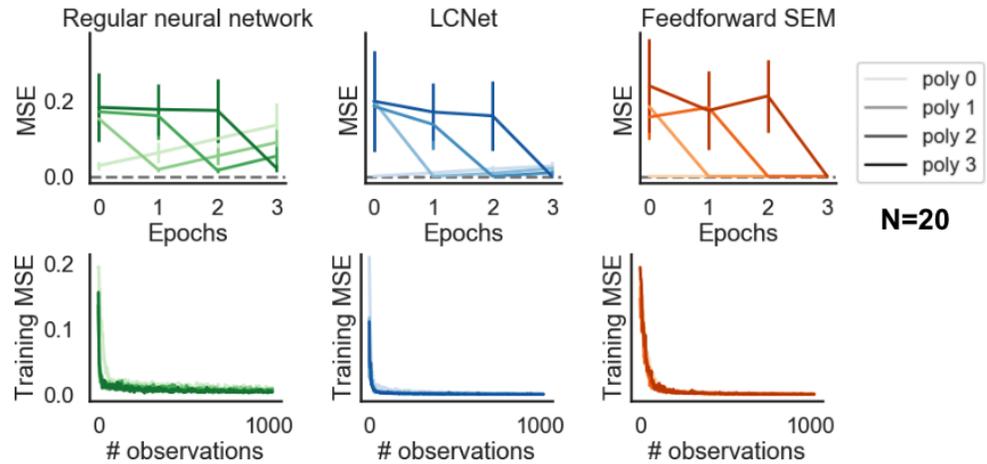
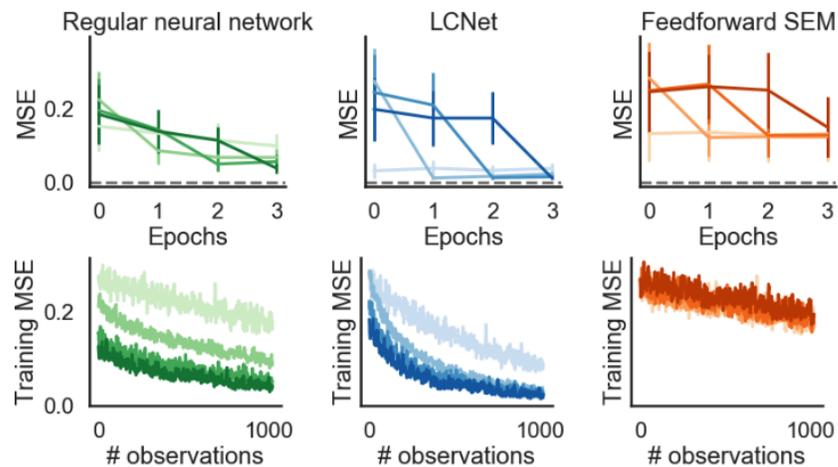

Figure S11

Here, we report the main analyses (see Figure 2h, 2i) when the learning rate (which is $1 \times 10^{-5}$ in the main paper) is changed to $1 \times 10^{-4}$ (a) and $1 \times 10^{-6}$ (b). The same pattern reported in the main paper qualitatively holds. Error bars indicate 3SE. $N = 20$ models per condition.